\documentclass[11pt]{extarticle}
\usepackage{geometry}                
\usepackage[parfill]{parskip}    
\usepackage{graphicx}
\usepackage{amssymb}
\usepackage{epstopdf}
\usepackage{mathrsfs}
\usepackage{stmaryrd}
\usepackage{bm}
\begin{document}

\parskip=6pt

\begin{center}
{\Large\bf  Hume's dictum as a guide to  ontology}\\

Adam Caulton (adam.caulton@philosophy.ox.ac.uk), 29 September 2024
\end{center}

\begin{abstract}
In this paper I aim to defend one version at least of Hume's dictum: roughly, the idea that possibility is determined  by ontology through something like independent variation. My defence is broadly pragmatic, in the sense that adherence to something like Hume's dictum delivers at least three benefits. The first benefit is that, through Hume's dictum, a physical theory's ontology delimits a range of possibilities, that I call \emph{kinematical possibilities}, which serves as a sufficiently permissive notion of possibility to sustain something like an intensional semantics for its claims, and a sufficiently demanding notion of supervenience to sustain  plausible claims of inter-theoretic reduction and theoretical equivalence. The second benefit is that Hume's dictum allows us to work backwards from a range of kinematical possibilities to an ontology. This is especially useful when aiming to glean an interpretation of a physical theory, since often we are more confident that we have arrived at the right space of possibilities than that we have arrived at the right ontology. The third benefit is that Hume's dictum---at least the version I aim to defend here---may be applied to physical theories more or less as we find them, and therefore we can practice something resembling ontology without having to force our theories into some Procrustean bed, such as a first-order language.
\end{abstract}

\tableofcontents

\section{Hume's dictum}\label{Hume}

In this paper I aim to defend one version at least of Hume's dictum: roughly, the idea that possibility is determined  by ontology. My defence is broadly pragmatic, in the sense that adherence to something like Hume's dictum delivers at least three benefits. The first benefit is that, through Hume's dictum, a physical theory's ontology delimits a range of possibilities, that I call \emph{kinematical possibilities}, which serves as a sufficiently permissive notion of possibility to sustain something like an intensional semantics for its claims, and a sufficiently demanding notion of supervenience to sustain  plausible claims of inter-theoretic reduction and theoretical equivalence. The second benefit is that Hume's dictum allows us to work backwards from a range of kinematical possibilities to an ontology. This is especially useful when aiming to glean an interpretation of a physical theory, since often we are more confident that we have arrived at the right space of possibilities than that we have arrived at the right ontology. The third benefit is that Hume's dictum---at least the version I aim to defend here---may be applied to physical theories more or less as we find them, and therefore we can practice something resembling ontology without having to force our theories into some Procrustean bed, such as a first-order language.

In the remainder of this section, I aim to get clearer on what I take Hume's dictum to say and what I take it not to say. In section \ref{Quine} is pay particular attention to a rival approach to ontology---that offered by Quine. This approach, I take it, is in many ways the received view, and therefore `the one to beat'. This discussion is continued in section \ref{Wallace}, in which I link up to themes from structural realism, particularly as recently expressed by Wallace (2022). In section \ref{Curiel} I outline the `upwards' use of Hume's dictum: the determination of a theory's kinematical possibilities from its ontology. Here I give especial attention to Curiel (2016), who seems to have been the first to notice the significance of kinematical possibility for the interpretation of physical theories. In section \ref{Weatherall} I turn specifically to the question of theoretical equivalence, with specific attention to the criterion advocated by Weatherall (e.g.~2016), and outline how attention to kinematical possibility might revise some existing judgments about when two theories are equivalent. In section \ref{Carnap} I outline the `downwards' use of Hume's dictum, and present what I believe to be examples of this implementation in the wild. 

The canonical statement of Hume's dictum appears in the \emph{Treatise} (1975 [1896], Book 1, Part III, \S6):
\begin{quote}
`There is no object, which implies the existence of any other if we consider these objects in themselves, and never look beyond the ideas which we form of them.
Such an inference [from cause to effect] would amount to knowledge, and would imply the absolute contradiction and impossibility of conceiving anything different.  But as all distinct ideas are separable, it is evident there can be no impossibility of that kind.'
\end{quote}
This statement is a crucial premise in Hume's attempt to eventually show that our concept of causation is founded on nothing but habit (i.e.~it is not a concept at all, but a mere {feeling} of expectation). In this passage, he is arguing that no \emph{demonstrable} connection between cause and effect exists. Thus the `objects' mentioned by Hume are more like what we would now call \emph{events}: local matters of fact. And the meaning is clear: there is no impossibility in an event we normally take to be a cause occurring and an event we normally take to be its effect failing to occur.

My interest here is not in doing full justice to Hume's words, but in extending a general idea. That idea, I take it, is that what is possible is determined by what there is, in the sense that all consistent combinations of existence/non-existence of distinct objects, or occurrence/non-occurrence of distinct events, give rise to a possibility, such that these possibilities so generated are in some sense as broad as it gets: i.e.~the space of possibilities is  ``ungappy''. Certainly, many modern formulations of Hume's dictum have been articulated and discussed (e.g. Wilson 2010, Russell \& Hawthorne 2018) which share this rough idea. The view also appears in the works of Armstrong (1986) and Skyrms  (1993) under the banner `combinatorialism', where it is linked to the early Wittgenstein's doctrine of logical atomism.

It has frequently been claimed that a counterexample to combinatorialism is provided by the fact that, for example, no object can be both red all over and green all over at the same time. More generally, it is widely thought to be subject to a `problem of determinables': for any determinable, why should one determinate property being instantiated by some object rule out the other determinates associated with the same determinable  being instantiated by the same object? It's not at all clear to me that this is something in need of explanation, but in any case it does seem clear that something like the combinatorial idea survives the mutual exclusion of determinates from the same determinable. For if we have, say, two objects $a$ and $b$ and two determinables $f$ and $g$, construed as functions from objects to the sets $F$ and $G$ of determinates, respectively, then some crisp sense of combinatorialism is still upheld if, despite the fact that e.g.~$f(a) = x$ excludes $f(a) = y$ for determinates $x\neq y$, nevertheless it is possible for $f(a)$ to take any value in $F$, while $g(a)$ takes any value in $G$, while $f(b)$ takes any value in $F$, while $g(b)$ takes any value in $G$, for a total of $|F|^{2}|G|^{2}$ possibilities.

It is this notion of combinatorialism that I wish here to defend. But many metaphysical ideas come under the adjective `Humean', so let me here explicitly spell out what I am \emph{not} defending.
\begin{itemize}
\item I do not wish to defend any kind of concept empiricism, like Hume's `copy theory' of perception. My defence of Hume's dictum is as a metaphysical thesis; I am not propounding any epistemological claim.
\item I do not wish to defend separability, in either its strong or weak forms (Butterfield 2011). Separability is a supervenience claim: namely that the intrinsic properties of any spacetime region supervene on the intrinsic properties of subregions in any partition of the original region, together with the spatiotemporal relations between those subregions. In fact, this supervenience claim would come out true if Hume's dictum were true, and the only external relations were spatiotemporal. But I have no case against external nonspatiotemporal relations.
\item I will not defend explicitly what Maudlin (2007, Ch.~2) called `Physical Statism': the view that `[a]ll facts about a world, including modal and nomological facts, are determined by its total physical state.' In particular, I do not wish to defend what is often described as a `Humean' view  about laws of nature. The view I wish to defend concerns kinematical possibility, not dynamical possibility (that is, possibility given the laws of motion). However, if it is correct that the kinematical possibilities are determined by a free variation of independent degrees of freedom determined by the theory's ontology, then it does seem there is no room for additional modal or nomological facts, for these would surely restrict the ``full'' gamut of possibilities in some way.
\end{itemize}
With these put to one side, let me turn to the view I am defending. As I said, this view is a kind of combinatorialism about possibility. The notion of possibility is question I call `kinematical', but only for lack of a better word. This notion of possibility is broader than dynamical, or physical, possibility, because it is a notion according to which the laws of nature might have been false. It is more restricted than metaphysical or logical possibility, because it is tied to an ontology---if you prefer, it is what it metaphysically possible, \emph{given} the ontology---but the existence of the ontology in question I take to be metaphysically or logically contingent.

One of my main contentions is that this notion of possibility is extremely useful within the philosophy of physics. And part of its extreme usefulness is how it is tied to ontology, via Hume's dictum. One example of its usefulness is in ruling out certain ontological claims, on the basis that, if those claims were true, one would otherwise have to posit cosmic conspiracies, primitive constraints, or other unacceptable brute facts. What makes these brute facts unacceptable is precisely that they rule out kinematic possibilities that, given the purported ontology and Hume's dictum, ought to be perfectly possible. A few examples of this kind of argument found in the literature are as follows.
\begin{itemize}
\item Maudlin (2007) on space. On the question whether distance relations between points or path lengths in space are fundamental, Maudlin argues that the former view requires us to
\begin{quote}
`brazenly postulate primitive constraints on the sets of distance relations that can be simultaneously instantiated between items, while substantivalists (who believe in the paths even when they are not occupied by material objects) need no such constraints.' (p.~88).
\end{quote}

\item Wallace (2014) on the Aharonov-Bohm effect. On Healey's proposed `loop ontology':
\begin{quote}
`[T]he ontology is very redundant: loops can be decomposed into smaller loops, and the real number assigned to the larger loop must be the sum of those assigned to its components. (If a region $R$ is simply connected, any loop can be decomposed into infinitesimal loops, and the $\mathbf{B}$ field of $R$ actually completely determines the values of all the loops in $R$.) Not only is this awkward, it is difficult to explain naturally \emph{except} by defining the values of each loop as the integral of some vector field around the loop.' (p.~6).
\end{quote}
This observation has been made by others, but I single Wallace out, because his own proposed ontology to understand the Aharonov-Bohm effect \emph{does} respect Hume's dictum---indeed I believe it is an example of an implementation of the `downwards' dictum, as I mention in section \ref{Carnap}. His ontology consists of just a scalar field $\rho$ (the magnitude of the erstwhile complex scalar field) and a 1-form $\mathcal{D}\theta$ (an ingenious  gauge-invariant construction out of the erstwhile complex scalar field's phase and the electromagnetic four-vector potential), and crucially there are no constraints placed on either.

\item Jacobs (2023) on comparativism. (Jacobs also addresses both of the examples above.) His general case against comparativism is encapsulated in the following quote:
\begin{quote}
`If the comparativist is to succeed, her theory must ``save the phenomena'', for even if there are \emph{in fact} only distances and mass ratios, the success of absolutist theories means that the world looks just \emph{as if} there are absolute positions and masses. In this paper, I argue that if comparativism is true, there is a sense in which it is a \emph{cosmic conspiracy} that the world looks just as if there are absolute quantities.' (p.~2).
\end{quote}
\end{itemize}

Thus we have out first implementation of Hume's dictum: in generating possibilities from an ontology. I return to this in section \ref{Curiel}. But another implementation I have in mind is that Hume's dictum might take us in the opposite direction, from possibilities to an ontology. So let me get into what I take to be the received view in contemporary analytic philosophy about how ontology is to be conducted.

\section{Quine's dictum}\label{Quine}

Quine (1948, 1960) offered a clear procedure for engaging in ontology. First, regiment your theory in a first-order language, in the most elegant way. The ontology suggested by your theory is then whatever the quantifiers range over. Thus Quine's dictum: to be is to be the value of a bound variable.

Let me first say something about what I think this approach gets right. The first is that existence claims are brought down to earth: if what you believe is your best theory quantifies over them, then you  believe they exist. This meshes with the use of Quine-Putnam indispensability arguments for various existence claims, and aligns with one aspect of the `Show me the theory!' objection deployed in the philosophy of physics. So, you think there is a way to save the phenomena successfully predicted by general relativity which makes no mention of spacetime points? Show me the theory! 

A related thing this approach gets right is, very broadly, the general idea of regimentation. It is only through regimentation of some kind or other that it is revealed to us what, exactly, a theory needs to posit for its claims to come out true, or at least approximately true. To return to the example just mentioned, perhaps the best argument for the existence of spacetime points is that no fully worked out classical spacetime theory with the same empirical successes as general relativity has been forthcoming which does not posit them, either primitively or derivatively.

Another thing that this approach gets right---and this perhaps something that is under-appreciated---is that regimentation into a first-order language offers a framework in which any claim made in the language of the theory is given a semantic value. I mean here not just a truth-value but something approximating what Carnap and Lewis called an \emph{intension}: a set of models of the theory's language in which the claim is true. I will get further into the import of this in section \ref{Curiel}, but here I will just briefly say that we are interested in understanding all sorts of claims one can make in a theory's framework, including claims that are compatible with the theory's central equations being false. To add to van Fraassen's  (1991) celebrated maxim that to interpret a theory is to specify what the world would be like were the theory to be true, I am suggesting that just as much as this we want some understanding of what the world would be like were the theory to be false.  And there are two ways for the theory to be false: it may have the ontology right, but have it wrong about how that ontology is arranged, or it may have failed even to get the ontology right. Now, it may be too much to ask to get a crisp understanding of the second kind of getting it wrong. But we can get a clear idea of the first kind of getting it wrong: we need only to make sense of possibilities in which the same ontology is arranged other than the theory claims. I believe this  becomes particularly salient when considering relations between theories, especially inter-theoretic reduction and theoretical equivalence.

Related to this is that what might be called `mere logical possibilities'---represented by  models of the first-order language---do obey, collectively, something like Hume's dictum as I conceive it. That is: in addition to  the fact that a model's domain may be \emph{any} non-empty set (and even the exclusion of the empty model is often seen as a convenience rather than a metaphysical postulate), the extension assigned by a model to a monadic predicate may be \emph{any} subset of the domain, the extension assigned to a binary predicate may be \emph{any} set of ordered pairs of objects from the domain, and so on. As a consequence, the supervenience of some new predicate on the predicates of the original language (given some defining theory, of course) is a particularly strong achievement---so strong, in fact, that by Beth's theorem it is equivalent to explicit definability.

However, there is something about Quine's approach, certainly to physical theories in particular, but perhaps more generally too, that isn't right. I hope to explain these shortcomings in the following few paragraphs. As a way in, consider first an apparent puzzle. Quinean regimentation calls for us to recast out favourite theory in some first-order language. But it is an elementary fact that any first-order theory cannot pick out a model except, at best, up to elementary equivalence. First of all, for realistic theories, which surely require infinite domains, this means that we cannot pick between models whose domains have differing infinite cardinalities. But put aside this issue and let us grant what is false, namely that we can pin a model down up to isomorphism. Precisely what differs among isomorphic models is what objects are in their domains! How then \emph{could} our theory, a mere set of sentences, tell us what ontology it claims for itself, when it can't choose between models disagreeing precisely on what is in their domains?

But this puzzle rests on a misunderstanding. Our theory has what Quine called an \emph{ideology} as well as ontology: that is, a collection of predicates. It is precisely the predicates that tell you what kind of objects your theory claims to exist (or not exist). If your theory's language contains the predicate `$x$ is green', and `$\exists x\ x\mbox{ is green}$' is a theorem of your theory, then your theory claims the existence of green things. The misunderstanding stems, I think, from the assumption that a regimentation into a first-order language must mean a regimentation into a entirely \emph{formal} language, `formal' in the sense that one has not (interpreted) predicates, like `$x$ is green', but only (uninterpreted) predicate \emph{symbols}, like $Px$. If this is what a Quinean regimentation is supposed to be, then indeed it is hard to see what a theory's ontology could be. But this, I suggest, is adequate reason to desist from this interpretation of Quine.

Where this puzzled objection really does seem to have bite, however, is when the predicates themselves seem to be in need of more interpretation than we have yet on offer. If my theory entails `$\exists x\ x \mbox{ is an electron}$', then fine, according to my theory there are electrons. But just what \emph{is} an electron? The issue is that, when it comes to more theoretical predicates, our only place to turn to get some grasp on what they mean is the theory itself. In that case, it seems we might as well use a formal predicate symbol `$Px$' in place of (e.g.)~`$x$ is an electron', and let the meaning  of `$Px$' be fixed, in as much as it can be, by the logical relations that sentences containing the symbol have, in conjunction with the theory, to all other sentences in the language. 

That brutally brief summary glosses over what is now almost a century of debate over the semantics of theoretical terms, which I do not wish to rehearse here. I will just point out that the debate continues, and what I take to be genuine progress has been made, even rather recently (I would single out Lutz (e.g. 2012) and Andreas 2021 in particular here.) Certainly, Quine had his own position on the matter, which appealed to the so-called `web of belief'. The upshot of this position for our current problem appears to be this: if you ask me what an electron is, and what it means to say that there are electrons, I will point you to the variety of sentences  connected, by way of logical inference and assumption of the theory's axioms, to that existence claim, and the variety of sentences connected to those sentences\ldots until we bottom out at the so-called `occasion sentences', and that is the end of the matter.

I will first say that this does not strike me as a realist position. The semantic holism obscures that this is essentially a verificationist proposal, albeit one liberated from the `two dogmas' of reductionism and the analytic/synthetic distinction. In any case, I want to point out that we \emph{can and do} often say more---much more!---in response to questions like, `What is an electron?' And what we can and do say in addition to the eventual bottoming out in `occasion sentences' involves appealing to the mathematical structures employed by the theory, and these appeals are genuinely enlightening as to the kinds of things our theories say exist.

\section{Predicate precisifications}\label{Wallace}

This is not an original observation. Recently,  Wallace (2022) has proposed a form of scientific realism he calls \emph{Maths-First Realism}, which he connects with structural realism of a kind that has been articulated  in the past couple of decades by many others (e.g.~Ladyman 1998, 2001, Saunders 2003, Ladyman \& Ross 2007, French \& Ladyman 2010, French 2017, McKenzie 2017). There are also connections to much earlier work in the so-called `semantic approach' to scientific theories, particularly the idea of state space semantics, pioneered by Beth (1960) and Bas van Fraassen (1967, 1970).

Quinean regimentations of the kind just discussed fall under what Wallace calls \emph{predicate precisifications}, and are tied to what he calls the `language-first' view of theories. Here is Wallace (2022, p.~363):
\begin{quote}
`[T]heories', on the language-first view, are predicate precisifications of mathematically-presented theories, and so (a) theories may have logically-inequivalent precisifications (leading to underdetermination) and (b) there may be no straightforward logical relation between the precisifications of an older and a newer theory, even when there is a mathematically-well-behaved instantiation relation between them. If the language-first view gave the correct account of scientific practice, then different precisifications are different theories (so that theory choice confronts under-determination) and theory change involves radical change of theoretical content, including of ontology.'
\end{quote}
I couldn't agree more.  Like Wallace, I agree that physical theories come with their own regimentation: namely, the mathematical formalism in which they are standardly presented. But Wallace appears to be more sceptical about ontology---that is, the \emph{practice} of ontology---connecting it to predicate precisifications, which bring the threat of under-determination and radical theory change. What I am suggesting is that ontology should begin and end with a consideration of a physical theory's mathematical formalism to which it is indigeneous. What is universal, or nearly universal, about physical theories is that they come with a space of possibilities, and this is our way in to doing ontology, thanks to Hume's dictum.

What I want to emphasise is not just that predicate precisifications give us too much, as it were. It's not just that predicate precisifications rest on decisions that seem more like stipulations than hypotheses, and so take us further away from a reliable picture of what is really `going on'. It is also that  engaging in predicate precisification causes us to lose something, or at least overlook the most important things our theories are telling us.  Allow me to run through a few examples to illustrate what I mean.

\emph{Example 1.} According to our best classical theory of the electromagnetic field, whatever it is, it is represented by the Faraday tensor field, which is a 2-form. Now, a 2-form is a very special kind of thing. First of all, it is what in differential geometry is called a `geometric object', namely an object that is invariant under any coordinate transformations. Secondly, the fact that it is a rank-$(0,2)$ tensor field and  anti-symmetric in its two indices means that it assigns scalars to unparameterised \emph{surfaces}. And it does so in a way that obeys simple additivity: i.e.~the scalar assigned to the surface $S_1 \cup S_2$ is just the sum of the scalars assigned to each of $S_1$ and $S_2$. This is what licences the `good myth' (in Lewis Carroll Epstein's sense) that magnetostatic fields are persisting field lines, permeating all of space, which intersect any given surface a determinate number of times. This is in some sense `all mathematics', since the various possible values for (say) the $\mathbf{E}$  field are not further interpreted (as the velocity of bits of aether, for example)---except perhaps that we \emph{do} assume that the manifold over which it is defined represents honest-to-goodness physical spacetime. But this mathematics in fact gives us a great deal of information about the form---the structure---of the electromagnetic field. And this is all without bringing `occasion sentences' into it---it is without even bringing charge currents into it.

\emph{Example 2.} Suppose our physical system has (say) 17 degrees of freedom. The number of degrees of freedom a system has is a remarkably robust fact. It is invariant under a switch from a Lagrangian to a Hamiltonian treatment, for example. It is even invariant under how the system's configurations space is coordinatised into generalised coordinates. But degrees of freedom are ontologically significant. Not only is the number of degrees of freedom of a system  crucial in predicting the thermodynamic behaviour of that system (e.g.~the difference between a monatomic and diatomic ideal gas); correctly identifying the number of system's degrees of freedom is also important in theory interpretation. For example, it allows us to correctly judge that in relativistic spacetimes a time-like worldline's four-velocity is not a genuine `motion through spacetime', whatever that was ever supposed to mean, but nothing but a \emph{direction in} spacetime. Why is that? It's because at any point along that worldline, the four-velocity has three, rather than four, degrees of freedom.

\emph{Example 3.} Quine's approach, and the idea that `to be is to be the value of a variable', misidentifies the sorts of things we think of as objects in scientific practice. Take for example a hydrogen atom. Now surely we can construct a reasonable toy first-order theory that quantifies over hydrogen atoms. And `reasonable' here means: we could come up with some axioms about hydrogen atoms that we take as true in some highly restricted regime, and use them to generate some  theorems that we take as true in that regime. But this toy theory could only be that: a toy theory. Hydrogen atoms are just not the kinds of things that could be the value of a first-order variable in any plausibly fundamental theory. That is to say: there \emph{is} no plausible predicate precisification of, say, the Standard Model of particle physics, in which things like hydrogen atoms come out as objects in the Quinean sense. There is not even a plausible predicate precisification of non-relativistic many-particle quantum mechanics in which hydrogen atoms come out as objects in the Quinean sense. And what I think this suggests is not that hydrogen atoms are not objects, or that the whole practice of ontology ought to be consigned to the dustbin of history. I think it suggests that predicate precisifications are just the wrong sort of things to look for when practising ontology in a naturalistic spirit.

\emph{Example 4.} In any kind of project of inter-theoretic reduction, we find that we need some way of, roughly speaking, constructing the big things (of the reduced theory) out of the small things (of the reducing theory). The resources available for  first-order languages are not up to this task. I wish to briefly discuss three ways it  has been suggested this might go in a first-order theory. In order of increasing ambition they are: mereology, Morita extensions and set theory.
\begin{itemize}
\item Mereology is favoured by some as a way of building big things out of smaller things, since it has at least some plausible claim to be an extension of logic worthy of the name (e.g.~Lewis 1991). This is important, since the degree to which our theory of construction fails to be logical is the degree to which our purported reduction fails to be to the would-be reducing theory, as opposed to the reducing theory \emph{plus} our theory of construction. But physical systems are just typically not constructed from other physical systems in anything like the way described by mereology---indeed the assumption that they are has been rightly dismissed as `the philosophy of A-level chemistry' (Ladyman \& Ross 2007, p.~24).
\item Morita extensions are the newest idea on the scene, and have been presented in a number of recent works by Barrett and Halvorson (see e.g.~Barrett \& Halvorson 2016). Morita extensions allow for constructions far more diverse than what would be allowed by mereology, and  which also have a plausible claim to be logical constructions. And one wonders whether Morita extensions might somewhat  mitigate the apparently arbitrary choices one makes in specifying a predicate precisfication, in the sense that the same ontology pops up, via Morita extensions, no matter which objects are taken as first-order primitives. But Morita extensions are still tied to first-order theories. Could we construct a hydrogen atom from certain states of the fields of the Standard Model \emph{via} Morita extension? It seems extremely doubtful. But even if it were possible, it would require casting our theories into first-order form.
\item Set theory was suggested by Lewis (1970) as a means to construct the entities of a reduced theory from those of a reducing theory. But set theory is at the same time too much and not enough. It is too much, since the full gamut of set-theoretic constructions---including those made out only of pure sets---makes the project of object construction for the reduced theory far too easy. It is not enough in the sense that learning that, for example, a box filled with gas is some unholy set of sets of sets of \ldots engenders more confusion that enlightenment.
\end{itemize}
Wallace (2022) has suggested that inter-theoretic reduction goes \emph{via} `structural instantiation': the mathematical structures of the reduced theory's models being instantiated or approximately instantiated by those of the reducing theory. (That idea can also be found in Morita extension, albeit restricted to first-order theories.) The key idea here is surely the familiar one of supervenience: the structures of the reduced theory's models is shown, via some representation theorem, to supervene on those of the reducing theory's models. The conception of ontology as tied to kinematical possibility meshes perfectly with this idea: for if we can say that the kinematical possibilities generated by the reduced theory's ontology supervene on the kinematical possibilities generated by the reducing theory's ontology, we can thereby say that the reduced theory's ontology is derivative on the reducing theory's ontology, for the reducing theory's ontology  generates, through something like recombination, all the possibilities for the reduced theory's ontology.

Another upshot of abandoning Quinean regimentation and predicate precisifications, I think, is that the Quinean dichotomy between ontology and ideology falls by the wayside. For example, take general relativity in its usual presentation in terms of models $\langle M, g, \Phi\rangle$, where $M$ is a four-dimensional smooth manifold, $g$ is a Lorentzian metric and $\Phi$ is some physical field, e.g.~the electromagnetic Faraday tensor. It seems undeniable that we should count the manifold points in $M$ as part of the theory's `ontology', but what about $g$ and $\Phi$? These are tensor fields, and so are functions defined on $M$, and so in that way appear like properties, and so perhaps on the `ideology' side of the Quinean dichotomy. But then again they are the subject of predication (such as when we describe $g$ as `Lorentzian'), and so ought to lie on the `ontology' side. Must this be settled? What cannot be denied is that $M$, $g$ and $\Phi$ all contribute to generating  possibilities, so let them be all count as ontology.

\section{Upwards Hume's dictum: from ontology to kinematical possibility}\label{Curiel}

As we saw in section \ref{Hume}, Hume's dictum allows us to pass from an ontology---in the looser sense just articulated---to a space of possibilities, through something like an independent variation of distinct degrees of freedom. This space of possibilities is vastly larger than those picked out by the theory's equations of motion. This, I claim, is an advantage. In the absence of any universal framework in which to treat physical theories, such as first-order logic would be under Quinean regimentation, we need some substitute for what such a universal framework provides.

The first thing such a framework provides is, as briefly mentioned in section \ref{Quine}, an intensional semantics for any claim made in the language of the theory. Whatever else intensions assigned to sentences do, they provide us with clear truth-conditions for those sentences in the most straightforward way possible: they present us with the set of possibilities in which the sentence is true. The importance of this lies in the fact that,  we wish to understand very many claims one might make in the framework of a theory. Specifications of certain states, initial or boundary conditions, general constraints over the form of forces or potentials, and so on---these are all claims that reach out of the narrow confines of a theory's space of dynamical solutions. The kinematical possibilities provide a wider, well-controlled space of possibilities which we can think of these claims as dividing.

This also relates to inter-theoretic reduction, in at least two ways. The first way is that what we want from an inter-theoretic reduction is not merely the entailment of the reduced theory by the reducing theory. That is, if you like, only one entailment relation of interest (though admittedly an important one!). What we also want is a translation scheme for any old claim we might wish to make in the framework of the reduced theory. When we say, in the framework of thermodynamics, that the entropy of the system is such-and-such, what am I saying about the system in terms of the microphysics? If we construe part of the project of inter-theoretic reduction as determining, for each kinematical possibility of the reduced theory, what is or are the corresponding kinematical possibility or possibilities of the reducing theory, then we have a general recipe for translating arbitrary `propositions' in the framework of the reduced theory into `propositions' in the framework of the reducing theory.

The second relation to inter-theoretic reduction, as briefly mentioned in section \ref{Wallace},  is supervenience. Demonstrations of supervenience are our best case for the ontology of one theory being reducible to the ontology of another. Recall that supervenience is a covariance claim: $A$ supervenes on $B$ iff, over some specified gamut of possibilities, any change in $A$ is accompanied by some change in $B$. But supervenience is too easy if the gamut of possibilities over which we vary is too narrow. For example, if we restrict to dynamical solutions in a deterministic theory, the future supervenes on the past; but this is bad evidence that the future is \emph{reducible} to the past. The fact that the future is not reducible to the past is shown by the fact that, in the true spirit of Hume's dictum, kinematical possibilities exist with the same past but different futures.

Despite what I have so far argued, it has to be admitted that there are many examples of theories whose `kinematical possibilities' as normally conceived are extremely restrictive from the point of view of Hume's dictum. I will mention two specific examples.
\begin{itemize}
\item \emph{Quantum field theories.} Consider any quantum field theory characterised as a theory of local field operators obeying the Weyl algebra (as discussed e.g.~in Ruetsche 2011, \S3.3.2).  In any irreducible representation of this algebra, any state is related to any other by a \emph{local} (that is, confined to some compact spatial region) transformation. Metaphysically, there is no motivation for this at all, since of course global transformations (which would send us between inequivalent representations) make just as much sense as local ones. However, there is an excellent technical motivation for this: namely that the symplectic product, which characterises the algebra, is well-defined only for such local transformations.

\item \emph{Classical spacetime theories.} In the Earman (1989) tradition, a classical spacetime theory is often associated with a class of `kinematically possible models', or KPMs, of the form $\langle M, A_1, \ldots A_m, D_1, \ldots, D_n\rangle$, where $M$ is a four-dimensional smooth manifold, each $A_i$ is an `absolute' geometrical object (contributing to `spacetime structure') and each $D_j$ is a dynamical geometrical object (contributing to `matter'). Now, across the KPMs, the absolute objects $A_i$ are assumed to either be constant or constant up to isometry, while the dynamical objects $D_j$ are free to vary in very much a Hume's dictum sort of way.

These absolute objects violate Hume's dictum, since they surely count among the theory's ontology, yet they don't vary across kinematical possibilities. But in this case, I suggest we conceive of this lack of variation as an artificial restriction of the KPMs: for we are only interested in KPMs for which the absolute objects take the form that we commonly take them to have. As soon as we wish to consider, e.g.~inter-theoretic reduction relations between these absolute structures, then we must consider them against the backdrop of free variation.

For example, if we wish to demonstrate that the ``full'' Newtonian spacetime structure $\langle M, t_a, h^{ab}, \xi^a\rangle$ is equivalent to $\langle M, t_a, h_{ab}\rangle$, we show that $t_a, h^{ab}$ and $\xi^a$ are inter-definable with $t_a$ and $h_{ab}$, and this means demonstrating supervenience over possibilities in which the usual constraints on these fields are lifted. 

However, there is another approach to this issue which is more radical. According to this approach, the spacetime structure does not comprise a cluster of independent degrees of freedom at all, but rather they are mere `codifications' of the behaviour of matter. Taking this approach seriously would mean conceiving of spacetime structure like the metric(s) and affine connection as supervenient on the dynamical matter fields---arising when and only when the matter fields behave in the appropriate way. In that case, absolute spacetime structure would not be an (independent) part of the theory's ontology at all, and would therefore not contribute to the generation of kinematical possibilities under Hume's dictum.

I suggest that the idea just articulated offers an illuminating construal of what defenders of the so-called `dynamical approach' to spacetime structure are up to---for example Brown \& Pooley (2006), who dub the Minkowski metric a `glorious non-entity'; or Myrvold (2019), who argues that Earman's (1989) famous symmetry principles, according to which a spacetime theory's dynamical and spacetime symmetries must match, are analytic. I take it as advantage of my general approach that the difference between the `dynamical' and `geometrical' approaches to spacetime structure may be cast as one about what is kinematically possible. Indeed, according to the standard of theoretical equivalence I essay below in section \ref{Weatherall}, the `same' spacetime theory cast separately under the dynamical and geometrical approaches are not equivalent, because the latter posits vastly more kinematical possibilities than the former.
\end{itemize}
Thus kinematical possibilities in theories as we find them may be apparently restricted, either due to technical considerations, or simply because of what we happen to be interested in at the present moment. I wish to dig my heels in at this point: if we wish to address questions of physical interpretation, or reduction or equivalence, and unless we intend to engage in some sort of eliminativist reductionism, then we must expand the received kinematical possibilities to the wider class of possibilities generated by free variation of the ontology we take the theory to posit.  

At this point, I wish to make contact with the work of Curiel (2016). Here is a particularly significant passage:
\begin{quote}
[K]inematical constraints are, in a precise sense, analytic: they are made true solely by the meanings of the terms \emph{in the context of the theory}. In that sense they are like L-sentences in a Carnapian framework (Carnap 1956). Unlike L-sentences, however, they have non-trivial semantic content, for the constraints they impose on physical system are non-trivial. Not all types of physical system will satisfy them, viz., those systems not appropriately represented by the theory.
\end{quote}
I think this is exactly right, and I endorse the connection to Carnapian $L$-rules. I would add an ontological twist to what Curiel says here, namely that `in the context of the theory' can be sharpened to, `in the context of theory being about the kind of entities that it is about'. And the precise way in which kinematical constraints are non-trivial is that the entities which necessitate them are contingent: they might not exist, or might not have existed. In other words: kinematical necessity is ontological necessity, it is metaphysical necessity, given the ontology. (I am not sure that Curiel would follow me here.)

 Curiel (2016) seems to have been the first to appreciate the interpretative significance of kinematical possibilities. But there are important differences between my position and Curiel's, both in the general and the particular. General first: Curiel talks in terms of kinematical \emph{constraints}, which are often (but not always) equations, and I talk in terms of kinematical \emph{possibilities}, which are represented by models of some kind. The chief difference is that, according to my view and properly understood, kinematical `constraints' are not constraints of any kind.

Turning to the particular, on the notion of `kinematical' and `dynamical' that I have outlined above, I cannot agree with most of Curiel's illustrative examples of kinematical constraints, although I agree with some. Allow me to go through each of them (found in \S2 of Curiel's paper):\footnote{Curiel further separates kinematical constraints into \emph{local} and \emph{global}, depending on whether the constraint can be attributed to individual states of the system. I will not comment on this classification.}

\begin{itemize}
\item `Hooke's constant $k$ has physical dimension $\frac{m}{t^2}$.'

This would be a reasonable example of a kinematical constraint in my sense, since there seems to be no good reason to suppose that it fails to hold in any kinematical possibility. Since Hooke's constant (not really a constant, of course, since it varies from system to system) precisely characterises linear restoring forces, it cannot fail to have dimensions of $[force][length]^{-1} = [mass][time]^{-2}$. One possible quibble might be that Hooke's constant is not even defined for systems not subject to a linear restoring force, so it would be wrong to assign it any dimension at all for such systems. But this really is just a quibble: we could simply add the clause, `if it is defined at all'. A separate, more serious quibble is that the dimensional identity  $[force][length]^{-1} = [mass][duration]^{-2}$ goes via a dimensional analogue of Newton's Second Law ($[force] = [mass][length][duration]^{-2}$), which Curiel takes to be dynamical, not kinematical (see below).  But we can sidestep this quibble by amending the original claim so that it mentions dimensions of force and length, rather than mass and duration.

\item `the shear-stress tensor is symmetric in Navier-Stokes theory, $\sigma_{ab} = \sigma_{(ab)}$.'

The symmetry of the Cauchy stress tensor is linked to the conservation of angular momentum and the absence of torques that may arise in complex fluids. The former is dynamical (since the conservation holds only ``on shell'') and the latter can fail, e.g.~for liquid crystals in the nematic phase (Xing \& Radzihovsky 2008). In each case, the conservation of angular momentum and the absence of such torques, the resulting symmetry of the Cauchy stress tensor ought to be viewed as dynamical. That is: there are kinematical possibilities in which it fails to be symmetric.

\item `Kepler's Harmonic Law, $\frac{a^3}{T^2} = M$.'

This law holds exactly for a Newtonian body subject to a central gravitational potential, and only approximately (even in the Newtonian theory) for orbiting bodies in the presence of other (sufficiently distant and sufficiently light) orbiting bodies, such as planets in our Solar System.  The fact that it holds only approximately shows that it cannot be a kinematical constraint in the sense of being kinematically necessary. Any claim to the effect that it is kinematically necessary for systems \emph{of a suitable type} (e.g.~Newtonian bodies subject to a central potential) is  illegitimate `smuggled' necessity, as described above, since there is sufficient similarity between orbiting bodies that obey it and those that don't obey it to block any ontological  distinctness claim.

\item  `stress-energy tensor is covariantly divergence-free in general relativity, $\nabla^n T_{na} = 0$.'

There is already something suspicious about this being a kinematical constraint, since in the case of flat  spacetime, this is a conservation law linked (via Noether's Theorem) to the invariance of the action under spacetime translations, which is unquestionably a dynamical constraint. What \emph{does} seem to be a good candidate for a kinematical constraint is the covariant divergence-freeness of the Einstein tensor, $\nabla^n G_{na} = 0$, since this follows from the definition of $G_{ab}$ and the Bianchi identities.\footnote{Are the Bianchi identities kinematical necessities? Some dispute has broken out about this. MORE} Now, the covariant divergence-freeness of the stress-energy tensor $T_{ab}$ does follow from this in conjunction with the Einstein field equations, which are surely dynamical. This would seem to understate the degree of necessity of $\nabla^n T_{na} = 0$, since the EFEs are merely sufficient but not necessary for it: that is to say, different dynamics would also lead to a covariantly divergence-free stress-energy tensor. Nevertheless, the covariant divergence-freeness of the stress-energy tensor follows (if it all) from the dynamics, so it is not kinematically necessary.

\item `the Heisenberg uncertainty principle, $\Delta x\Delta p \geq \frac{1}{2}\hbar$.'

This principle follows fairly straightforwardly from the `Heisenberg relation' between $x$ and $p$, namely $[x, p] = i\hbar$. This is an algebraic relation between quantities defined over the full Hilbert space of the system, and independent of any choice of Hamiltonian. It seems therefore to be a perfect candidate for being kinematically necessary in our sense.

\end{itemize}

Finally, Curiel gives as `the canonical example' of a dynamical equation of motion Newton's Second Law: `a Newtonian body accelerates in direct, fixed proportion to the net total force applied to it, the ratio of the acceleration to the total force being the kinematic quantity known as the body's inertial mass.' But I am not sure that this example is so canonical. This might seem puzzling, if not downright muddled: \emph{surely} if anything is a dynamical equation of motion, then Newton's Second Law is!

Let me clarify: particular \emph{instances} of Newton's Second Law, in which the form of the various forces have been explicitly specified---for instance, the equation of motion for a body in a central gravitational field (which implements the Universal Law of Gravitation), or for a body subject to a linear restoring force (which implements Hooke's Law)---are unquestionably dynamical equations of motion. But the Second Law \emph{without any specification whatever} of the form of the forces is not an equation of motion at all. Indeed, we cannot even say what the order of the would-be differential equation is, until we add further constraints, e.g.~that the forces must be functions only of positions and velocities.

So what then is Newton's Second Law, in its full abstraction? A kinematical constraint? I think the answer anyone should give to this, in line with the general position defended in this paper, hangs on whether or not they countenance forces as independent quantities, i.e.~new degrees of freedom. If forces are truly independent quantities, then one one way to make this precise they are  vector quantities that are relational: i.e.~forces are always of the form, \emph{the force of type $\alpha$ impressed by $a$ on $b$ is $\mathbf{f}^{(\alpha)}_{xy}$}, where $\mathbf{f}^{(\alpha)}_{ab}$ is some 3-vector, the components of which have dimensions of force. And they are truly independent quantities iff  it is possible (`kinematically possible') for the vector sum of them, say $\sum_{z, \alpha} \mathbf{f}^{(\alpha)}_{za}$ acting on any given body $a$, to take values other than $m_a\ddot{\mathbf{x}}_a$, the product of $a$'s mass and acceleration. (Wigner (1954), for example, entertains possibilities in which $\mathbf{F} = m\mathbf{v}$.)

If forces are not truly independent quantities, then they would have to be derivative: something like a convenient codification of the behaviour of bodies which exhibit certain trajectories. This is what becomes of forces in Sneed's (1971) analysis, for example: they are a fictional adornment on particle trajectories precisely in those cases where such an adornment would be compatible with Newton's Second Law. Thus on this view we can truly say: necessarily, if there are forces, then Newton's Second Law holds of them. This would make the Second Law a kinematical necessity in precisely my sense.\footnote{Of course, forces could still be derivative, and yet their existence hang on conditions more or less strict than simply the holding of Newton's Second Law. Perhaps we could tighten the conditions, so that all three of Newton's Laws must hold, or relax the conditions, so as to allow for forces in relativistic regimes.}

It is simply not clear to me which of these views is correct, and the topic has been extensively discussed  before, e.g.~Earman \& Friedman (1973), Nagel (1979, pp. 174-196). It is surely of great metaphysical importance which view is correct. But whichever one it is, we have a clear way of articulating what each view is committed to, in terms of what is kinematically possible.

Consider another tricky case of great importance: the Schr\"odinger equation. I mean here not, for example, the non-relativistic  Schr\"odinger equation for a particle in three dimensions, i.e.
\begin{equation}
i\hbar\frac{\partial \psi}{\partial t} = \frac{\hbar^2}{2m}\nabla^2\psi + V\psi
\end{equation}
(perhaps with some specification of the form of $V$),
but the much more abstract claim that the temporal evolution of the quantum state is determined by \emph{some} Hamiltonian, that is, some self-adjoint linear operator defined on the Hilbert space:
\begin{equation}
i\hbar\frac{\textrm{d}}{\textrm{d} t}|\psi\rangle = H|\psi\rangle\ .
\end{equation}
This equation follows purely, \emph{via} Stone's Theorem, from the assumption that the `trajectory' $|\psi(t)\rangle$ is generated by a continuous one-parameter family of unitaries, i.e.~$|\psi(t)\rangle = U(t)|\psi(0)\rangle$. This amounts to the assumption that, no matter the initial quantum state, it evolves continuously and so as to preserve the norm of the state-vector (in other words, so as to conserve probability).

I don't know what to say about this case, except to make a few scattered observations. First, the Schr\"odinger equation, even in this abstract form, is violated for `open' quantum systems, in which the dynamical evolution is permitted to be non-unitary (for example in the Quantum Fokker-Planck equation). Second, this abstract form of the Schr\"odinger equation is a common feature of all quantum theories in which the system in question is taken as closed.

To close off this section, it may help to provide some additions  to Curiel's list of conditions which are more plausibly kinematically necessary in my sense.
\begin{itemize}
\item The metric tensor $g_{ab}$ in general relativity is a non-degenerate, symmetric, rank-2 covariant tensor field, with signature $(+, -, -, -)$.
\item If the electromagnetic one-form $A$ is taken as primitive, then the Faraday tensor is closed, $dF = 0$.
\item The stress-energy tensor in general relativity is symmetric, $T_{ab} = T_{(ab)}$.
\end{itemize}
It may seem a stretch to call these conditions \emph{analytic} in any plausible sense, but I hope it is obvious in what sense they are \emph{necessary}, given what physical entities the corresponding mathematical objects are supposed to represent.

\section{Theoretical equivalence as kinematical and dynamical equivalence}\label{Weatherall}

I wish now to get into some more detail regarding theoretical equivalence. The topic of theoretical equivalence has received a lot of attention in recent years (for an excellent overview, see Weatherall 2019a, 2019b). A  criterion for theoretical equivalence that has gained significant attention, if not popularity, over the last decade or so, is due to Weatherall. It puts categorical equivalence at the centre of the conception of theoretical equivalence. More specifically, two theories $T_1$ and $T_2$, construed as categories of models, are equivalent iff the following conditions hold.
\begin{itemize}
\item[(i)] $T_1$ and $T_2$ are equivalent as categories. That is, there is a functor $F$ from $T_1$'s objects and morphisms to $T_2$'s objects and morphisms that is full and faithful (i.e.~for any two objects $a, b$ of $T_1$, $F$ is a bijection $F: \mbox{Hom}_{T_1}(a, b) \to \mbox{Hom}_{T_2}(F(a), F(b))$) and essentially surjective (i.e.~for any object $z$ of $T_2$ there is some object $a$ of $T_1$ such that $F(a)$ is isomorphic, by the lights of $T_2$, to $z$; there is no need to demand essential injectivity, since this follows from $F$'s being full and faithful). 
\item[(ii)] The functor $F$ that realises the categorical equivalence in (i) preserves empirical content.
\end{itemize}
An alternative standard, advocated by Barrett \& Halvorson (2016) and applying to first-order theories, is Morita equivalence, according to which (very roughly) two theories are equivalent iff they admit of a common Morita extension. As Barrett \& Halvorson show,  this is a strictly more demanding standard of theoretical equivalence than categorical equivalence, if we make the comparison possible by transmogrifying any first-order theory into a category by taking its models as objects and its elementary embdeddings as morphisms (thus `isomorphism' is in fact elementary equivalence). But I wish to focus on categorical equivalence.

Categorical equivalence has come in for some criticism as an acceptable standard of theoretical equivalence (Barrett \& Halvorson 2016, Hudetz 2019, Weatherall 2021); all criticisms to the effect that it is too easy to fulfil.
Something that bears emphasising about the implementation of categorical equivalence as a standard of theoretical equivalence in practice (e.g.~Rosenstock \emph{at al} 2015, Rosenstock \& Weatherall 2016, Weatherall 2016, 2020), and the received conception of physical theories as a category of models more generally, is that the theories in question are always conceived such that their objects are what are usually dubbed their \emph{dynamically possible models}. On this conception, we lose something I think is crucial: namely, the idea of the theory as ruling some possibilities in \emph{and other possibilities out}.

What I have in mind is that the inner workings of theory 1 may be able to make sense of kinematical possibilities that theory 2 cannot, and this difference is invisible to the above standard of equivalence if those kinematical possibilities are ruled out by the dynamical equations. Yet that difference is significant: for the fact that the inner workings of theory 1 can make sense of these possibilities, even though they are ultimately ruled out, while theory 2 cannot even make sense of them, suggests that the theories have some deep disagreement which should block their equivalence. I want to suggest that this disagreement is over \emph{ontology}: theory 1's ontology is richer than theory 2's, as evidenced by the fact that it admits kinematical possibilities that theory 2 does not. And two theories should not be counted equivalent if they disagree over ontology.

On one way of thinking about it, theoretical equivalence is entirely straightforward: two theories are equivalent precisely when they express the same proposition (Coffey 2014). The difficult question is how to identify when it is that two theories express the same proposition, when all we have to go on is our theories' empirical content and mathematical formalisms. What I mean to suggest is that progress on this question can be made by reference to kinematical possibilities. For if two theories may be said to share the same kinematical possibilities (or, more precisely, if their mathematical structures are such that they each admit of an interpretation in which they share the same kinematical possibilities), then they may be said to share a common semantic framework in which questions of equivalence are well posed.

This suggests a simple fix to Weatherall's proposed standard of theoretical equivalence. The first part of the fix is in the conception of theories as categories in general: the new idea would be to characterise a theory as a \emph{pair} of categories $\langle T_K, T_D\rangle$, the second a sub-category of the first, where the first category delimits the kinematically possible models, while the second category specifies the dynamically possible models. The second part of the fix is to strengthen the standard of theoretical equivalence so that we demand categorical equivalence between both the theories' categories of kinematically possible models and their categories of dynamically possible models. (I also intend to preserve the empirical equivalence condition, but this will not feature greatly in what follows.)

It is worth running through a handful of existing theoretical equivalence claims in the literature, to see what effect this fix has on the plausibility of those adjudications. Each case deserves far more attention than I have space to give here, but I hope nevertheless to convey a general flavour of the main idea.
\begin{itemize}
\item \emph{General relativity and Einstein algebras} (Rosenstock \emph{at al} 2015). This example presents an immediate problem for my proposed fix, since the theory described as `general relativity' by Rosenstock \emph{et al} is in fact \emph{already} a category of kinematically possible models, of sorts. The models are all of the form $\langle M, g\rangle$, where $M$ is a smooth, four-dimensional manifold and $g$ is a smooth Lorentzian metric, and the morphisms are isometries. No other constraints are placed on $g$---certainly no dynamical equations. Likewise, the theory of `Einstein algebras' is taken to be a category of models of the form $\langle A, \hat{g}\rangle$, where $A$ is a `smooth' Einstein algebra, as traditionally conceived,  $\hat{g}$ is a distinguished element, also a Lorentzian metric (albeit characterised as such algebraically), and the morphisms are algebraic homomorphisms which map distinguished elements to distinguished elements.  No other constraints are placed on $\hat{g}$.

However, this seemingly mad characterisation of `general relativity' has reasonable justification: for in a conception of that theory in which the models take the form $\langle M, g, T\rangle$, where $T$ is the stress-energy tensor,  any kinematical possibility for $\langle M, g\rangle$ is associated with exactly one dynamical solution $\langle M, g, T\rangle$ to the Einstein field equations. Thus Rosenstock \emph{et al} are surreptitiously taking $T$ to be determined \emph{via} the EFEs, and taking kinematical possibilities $\langle M, g\rangle$ as proxies for dynamical possibilities $\langle M, g, T\rangle$.

I have a few objections to this approach, the most obvious being that it rides roughshod over the importance of kinematical possibility as apart from dynamical possibility. I also object to the fact that this KPM-goes-proxy-for-a-DPM strategy requires thinking of the matter fields in general relativity as completely specified by the stress-energy tensor $T$. This is a mistake, for one cannot typically reconstruct matter fields, such as the Faraday tensor, from its associated stress-energy tensor. Moreover, the matter fields will obey their \emph{own} dynamical equations, and aside from the fact that these equations cannot be stated just in terms of $T$, they may in fact rule out some models $\langle M, g\rangle$.

But all of these objections may be overcome. Let us identify the \emph{kinematical possibilities} of general relativity instead with the category of models of the form $\langle M, g, \Phi\rangle$, where $\Phi$ now appears as a placeholder for whatever matter fields one wishes to treat, and which contribute to the stress-energy tensor, and let the morphisms be $(g, \Phi)$-preserving diffeomorphisms. Likewise we may identify the kinematical possibilities of the Einstein algebras with the category of models of the form $\langle A, \hat{g}, \hat{\Phi}\rangle$ in the obvious way. It is fairly clear that Rosenstock \emph{et al}'s original categorical equivalence theorem carries over with the mildest tweaks to these two categories of kinematical possibilities. Moreover, we can use the equivalence functor between kinematical possibilities to ensure that any sub-category of dynamical possibilities in general relativity (demarcated by laying down the EFEs, perhaps some energy conditions, and non-gravitational equations of motion for $\Phi$) is equivalent to some subcategory of dynamical possibilities for the Einstein algebras. Thus the theoretical equivalence can be (made to be) preserved under my proposed fix.

\item\emph{Generalised holonomy maps on a connected manifold and principal connections on that manifold} (Rosenstock \& Weatherall 2016). A `generalised holonomy map' is taken to be a function $H$, which given each point $x$ of the spacetime manifold $M$, yields a map $L_x$ from the smooth loops passing through $x$ into some structure group $G$, satisfying certain conditions. Thus we obtain a `holonomy model' $\langle M, H\rangle$, which are the objects of our first category. The objects of our second category are models of the form $\langle P, M, \pi, \Gamma\rangle$, where $P\to^\pi M$ is a principal $G$-bundle, and $\Gamma$ is a connection on $P$. The categorical equivalence relies on a representation and reconstruction theorem  due to Barrett (1991).
 
 In this case the categories of kinematical possibilities associated with each theory are fairly obvious, and it is obvious that they don't match.  For the generalised holonomy maps are constrained to obey the condition $H(\gamma_1\circ\gamma_2) = H(\gamma_1)H(\gamma_2)$, for any two composable loops $\gamma_1, \gamma_2$, and there is no  reason at all to preserve this constraint for the  kinematical possibilities. Thus the theoretical equivalence claim fails.

\item \emph{Classical electromagnetism with the Faraday tensor and with the four-vector potential} (Weatherall 2016). The first theory, $EM_1$, involves models of the type $\langle M, \eta_{ab}, F_{ab}\rangle$, with smooth four-dimensional manifold $M$, Minkowski metric $\eta_{ab}$ and Faraday tensor (2-form) $F_{ab}$. The second theory, $EM'_2$, involves models of the type $\langle M, \eta_{ab}, [A_{a}]\rangle$, where $[A_a]$ is a gauge orbit of four-vector potentials (1-forms) $A_a$, where $A'_a\in[A_a]$ iff $A'_a = A_a + d_a\chi$ for some smooth scalar field $\chi$. Morphisms in each case are structure-preserving maps in the obvious way.

So these are the categories of dynamical possibilities. Leaving aside alternative metrics, the kinematical possibilities for $EM_1$, I suggest, should include any possible configuration for a 2-form $F_{ab}$---crucially, configurations in which $F_{ab}$ fails to be closed, i.e.~$d_a F_{bc} \neq \mathbf{0}$. Likewise, kinematical possibilities for $EM'_2$ should include any possible configuration for (an equivalence class of) 1-forms $A_a$. Now, since the claimed categorical equivalence goes via the identification $F_{ab} = d_a A_b$, which constrains $F_{ab}$ to be closed, there are kinematical possibilities for $EM_1$ which are not kinematical possibilities for $EM'_2$, and thus the claimed theoretical equivalence fails.

\item \emph{Newtonian gravitation and geometrised Newtonian gravitation} (Weatherall 2016). The first theory, $NG$, involves models of the type $\langle M, t_a, h^{ab}, \nabla, \varphi, \rho\rangle$, where $M$ is a smooth manifold, $t_a$ and $h^{ab}$ are compatible temporal and spatial metrics, respectively, $\nabla$ is a compatible, flat affine connection, and $\varphi$  and $\rho$ are scalar fields, representing respectively the gravitational potential and  mass density. The second theory, $GNG$, involves models of the type  $\langle M, t_a, h^{ab}, \tilde{\nabla}, \rho\rangle$, where now $\tilde{\nabla}$ is a compatible affine connection which may have curvature. The morphisms of $GNG$ are straightforwardly associated with structure-preserving diffeomorphisms, but the morphisms of $NG$ are not so obvious. Weatherall's key claim is that, if we allow the morphisms to include so-called `Trautman symmetries', which connect models of $NG$ giving rise to the same free-fall trajectories of test bodies, thus specifying the theory $NG_2$, we obtain an equivalence of categories.

Do the kinematical possibilities of these two theories match? Let us leave aside possibilities in which the various compatibility constraints fail, or in which $h^{ab}$ fails to be Euclidean. (I believe these \emph{are} kinematical possibilities, but considering them will not help us here, since this is absolute spacetime structure common to both theories.) As regards $NG_2$, presumably we should relax the Poisson equation, relating the gravitational potential $\varphi$ to the mass density $\rho$. But this is not the only dynamical equation implemented in $NG_2$: we also have  the equation specifying free-fall trajectories, $\xi^n\nabla_n\xi^a = -\nabla^a\varphi$. Let us for the moment entertain that we are able to relax this too, independently of the Poisson equation. (Although it must be remarked that the models as given have no room for free-fall trajectories as an independent degree of freedom. We might consider adding them as an additional element.) Crucial for us is that there appear to be kinematical possibilities in which the Poisson equation is obeyed, and yet the equation of free-fall is violated.

On the side of $GNG$, we may relax the `geometrised Poisson equation', which relates the curvature of $\tilde{\nabla}$ to the mass density field $\rho$ and the temporal metric $t_a$. But are we at liberty \emph{independently} to switch on or off the equation of free-fall, in this case $\xi^n\tilde{\nabla}_n\xi^a = \mathbf{0}$? It seems not. For, just as the geodesic principle is a theorem of general relativity (Geroch \& Jang 1975), Weatherall (2011) proved the same for $GNG$. Thus, any kinematical possibility for $GNG$ in which the geometrised Poisson equation is obeyed---along with certain energy conditions---is also one in which the equation of free-fall is obeyed.

So it appears we  have a mismatch of kinematical possibilities. 
But not so fast. For we may suspect that a Geroch-Jang-like theorem is available for $NG_2$ too. And indeed, a moment's reflection reveals that there must be such a theorem. All we need to do is use the equivalence of categories between $NG_2$ and $GNG$ to mirror, in $NG_2$, the ingenious worldtube constructions in Weatherall's (2011) proof of geodesic theorem in $GNG$. Thus, in $NG_2$, just as in $GNG$, the equation of free-fall comes as a theorem of the (geometrised or ungeometrised) Poisson equation and suitable energy conditions.

Now, given that the kinematical possibilities in each theory involve  relaxing only the (geometrised or ungeometrised) Poisson equation, and/or the energy conditions required for Weatherall's Geroch-Jang-style theorem, it would appear that we can easily obtain an equivalence of categories for the kinematical possibilities too. For the nature of the equivalence between the dynamical possibilities hangs on a decomposition of the Newton-Cartan connection $\tilde{\nabla}$ into a flat connection term and term attributable to the gravitational potential, and this decomposition can be stipulated to remain in the absence of the Poisson equation and energy conditions.\footnote{But not so fast again. For $GNG$ as usually treated comes with constraints on the form of the dynamical connection $\tilde{\nabla}$, in particular  that it is `rotationally flat' ($R^{ab}_{cd} = \mathbf{0}$) and may be decomposable as $\tilde{\nabla} = (\nabla, -t_b t_c \nabla^a\varphi)$ for some scalar field $\varphi$. If these constraints are dropped as kinematical constraints, then we will find many kinematical possibilities for $GNG$ for which there is no corresponding kinematical possibility for $NG$. I am very grateful to Eleanor March for clarity on this.} Thus, the theoretical equivalence claim can be upheld.

\end{itemize}
Allow me now to take stock of these examples. In the second and third cases, we found a failure of theoretical equivalence, and this can be traced to an inequivalence of ontology. In the second case, an ontology of loops and their holonomies is not equivalent to an ontology of connections, even allowing for  gauge-equivalence. In the third case, an ontology including a 2-form is not equivalent to an ontology including some associated 1-form, even allowing for gauge-equivalence. The first and fourth cases are more intriguing, since we do have equivalences, despite apparently very different ontologies (spacetime points vs.~possible scalar fields; a flat affine connection and a gravitational potential vs.~a curved affine connection). But in both of these cases, we see the success of supervenience over the widest reasonable possibilities, in both directions. And so, even though we  appear to have a disagreement between two theories in a pair over which entities are taken as fundamental or primitive, each theory in the pair is able to reconstruct the entities of the other theory.

to conclude this section, I will just advertise that March (2023) has  addressed the question of theoretical equivalence for Maxwell gravitation and Newton-Cartan theory, also with explicit consideration of each theory's mere kinematical possibilities. In this case, unobvious stipulations have to be made about kinematical possibilities, but not unnatural stipulations are possible according to which we obtain theoretical equivalence in my reformed sense. March (2024) also investigates theoretical equivalence claims more generally with kinematical possibilities taken into account.

\section{Downwards Hume's dictum: from kinematical possibility to ontology}\label{Carnap}

Finally, we come to what I think is the real payoff of Hume's dictum: implementing to work backwards from a space of kinematical possibilities to an ontology. The broad idea is that the ontology associated with a space of kinematical possibilities is what can serve as a supervenience base for those possibilities. The reason I take this to be such a payoff is that the `right' space of kinematical possibilities can often be epistemologically prior to the `right' ontology.
Frequently, that can be because we have started out with a naive ontology, used Hume's dictum to generate a naive space of kinematical possibilities, and then revised that space of kinematical possibilities by incorporation of constraints or consideration of symmetries. The classic example is of course gauge theories, in which our `naive' ontology overcounts the degrees of freedom of our system.

It bears emphasising that by `symmetries' I mean \emph{dynamical} symmetries, which pertain to the much more restricted class of dynamical possibilities. But `symmetry-to-reality' inferences (Dasgupta 2016) lead us to reduce the degrees of freedom we take our system to have, and thus reform the space of kinematical possibilities.

One of the most successful implementations of the `downwards' Hume dictum I know is Wallace's (2014) treatment of the Aharonov-Bohm effect. Starting with the naive ontology of a complex scalar field and four-vector potential, and then accounting for the gauge symmetry, and focussing solely on gauge invariant quantities, Wallace comes to the conclusion that the right ontology is a real scalar field and 1-form which is some amalgam of aspect of the original complex scalar field (namely its phase) and the four-vector potential. This solution perfectly eliminates gauge freedom without placing any unexplainable constraints on the new ontology.

Another implementation of the `downwards' Hume dictum can be made in the case of many-particle quantum mechanics of `equivalent' particles. In that case, the global quantum state is constrained to be either totally symmetric or totally anti-symmetric under a permutation of factor Hilbert space labels. According to the popular interpretation of this fact, all particles of the same species take the same (highly statistically mixed) state. There are many objections one might level against this interpretation (see Caulton 2024), but one is a simple appeal to Hume's dictum: that all particles of the same species must take the same mixed state is an unacceptable necessary connection. Is an alternative interpretation possible?

There is. For, in the joint Hilbert space, one may identify subspaces which possess, along with their associated algebra of observables, a tensor product structure, just as for `distinguishable' particles. That is to say: such subspaces may be decomposed into independently varying degrees of freedom. (The independent variation in this case is represented precisely by the tensor product structure.) According to this `anti-factorist' interpretation, so-called `indistinguishable' particles become perfectly distinguishable, in fact by monadic properties.

But these two are not the only examples of a `downwards' implementation of Hume's dictum. I wish to discuss four more examples, from philosophy, physics and two from general science. For reasons of space I must be brief.
\begin{itemize}
\item \emph{Carnap's quasi-analysis.} In the \emph{Aufbau} (Carnap 1969 [1928]), Carnap embarks on a heroic project of logical construction, in which all of the objects of science are to be constructed, in the end, from `elementary experiences'---total temporal cross sections of first-personal phenomenological experience---and a single relation between them: time-directed recollected similarity. Early on in the project of construction, Carnap constructs the various `quality classes': what from outside the system we would ordinarily think of as patches of colour in the visual field, or single tones in auditory experience.

The construction of quality classes from elementary experiences is a procedure than Carnap calls \emph{quasi-analysis}: `analysis', because it seems we are breaking the elementary experiences down into their basic components, from which everything else will be constructed; `quasi-', because, from the perspective of the system, elementary experiences \emph{are} already the basic components. I will not go into detail about quasi-analysis (see Pincock 2009), except to point out two themes. (i) quasi-analysis is an example of applying Hume's dictum `downwards': for the elementary experiences are global phenomenological states, and the products of quasi-analysis are an ontology. (ii) the success of quasi-analysis---in the sense of managing to derive, as the products of quasi-analysis, all of the basic components which we pre-theoretically wish to derive---hangs on the elementary experiences being sufficiently rich in variation. The failure of the elementary experiences to be sufficiently rich leads to what Goodman (1977) dubbed the `problem of imperfect community'. 

\item \emph{Einstein algebras.} The equivalence of modern differential geometry, as usually conceived in terms of tensor fields on smooth manifolds, and the theory of Einstein algebras, has already been discussed. But a significant fact about Einstein algebras is that their elements---tensor fields treated as primitive objects---must be understood as \emph{merely possible} tensor field configurations. In that sense, a single Einstein algebra (with no distinguished element) is something like a space of kinematical possibilities for a single manifold. And indeed, it is because this space is so incredibly richly varied and structured that one can construct manifold points---thus giving an Einstein algebras the same diffeomorphism-induced symmetries as manifolds in familiar modern differential geometry (Rynasiewicz 1992). 

This construction of manifold points is another example of a `downwards' Hume dictum, for a crucial condition on the reconstruction of the manifold being possible is that  the Einstein algebra is sufficiently rich. Indeed the reconstruction may fail if the Einstein algebra remains smooth while the associated manifold is non-Hausdorff (Wu \& Weatherall 2023). It fails in this case precisely because the non-Hausdorff manifold possesses degrees of freedom that are invisible from the perspective of the algebra of \emph{smooth} scalar fields, that is from the perspective of a restricted space of possibilities.

It is also worth commenting that the Einstein algebra/manifold equivalence has received two interpretations in the literature. Here is Rosenstock \emph{et al} (2015, \S 5):
\begin{quote}
Insofar as one wants to associate these two formalisms with ``relationist'' and ``substantivalist" approaches to spacetime, it seems that we have a kind of equivalence between different metaphysical views about spatiotemporal structure.
\end{quote}
Thus the equivalence signals something like an equivalence between traditional metaphysical rivals. By contrast, here is Rynasiewicz (1992, p.~588):
\begin{quote}
`[T]he program of Leibniz [Einstein] algebras is not a \emph{tertium quid} to the traditional substantival-relational debate, as Earman understands that debate, but is instead a recasting of substantivalism in different mathematical form.'
\end{quote}
Thus the equivalence signals that the theory of Einstein algebras is substantivalism in sheep's clothing. From the point of view of Hume's dictum, we must side with Rynasiewicz. For what differs between the standard theory of manifolds and tensor fields on the one hand, and the theory of Einstein algebras on the other hand, is not a difference in ontology, but merely a difference in what is taken as primitive.

\item \emph{Multi-dimensional scaling.} MDS is often presented as a technique in data visualisation, which has found a wide variety of applications across a number of different sciences, but especially psychophysics and neuroscience (Mead 1992, Wickelmaier 2003). The basic idea is fairly easy to state. One begins with a collection of `items' and some metric of similarity defined on them, typically in the form of a `dissimilarity matrix'. The name of the game is then to embed these items in some high-dimensional space, often assumed to have a Euclidean metric, such that the distances between the items represents their degree of dissimilarity.

In fact, one typically starts off embedding the items in some very high-dimensional space, and attempts to drive the number of dimensions down without doing too much violence to the original dissimilarity scores. Often one finds an optimal number of dimensions in which the items may be embedded with only slight tweaks to these dissimilarity scores. The resulting space's dimensions are then interpreted as the collection of items' degrees of freedom: that is, the dimensions over which the items may be interpreted as varying.

The connection of this technique to our theme is clear: for the `items' we begin with are typically highly structured, like states or possibilities. Then, through some structure of similarity defined on these items we derive degrees of freedom, on the assumption that the variation between these items is generated by an independent variation between the degrees of freedom.

\item \emph{Principal components analysis.} Like MDS, PCA is a technique used to find `latent variables' from structured data, but its applications are far wider in the sciences (Abdi \& Williams 2010). In this case, data points are plotted in some inner product space defined by manifest variables. Through analysis of these data points, principal components are identified as those directions in the  space along which the spread in the data points is the greatest. Also like MDS, PCA often leads to dimensional reduction. In this case, correlations between the manifest variables are explained away by a shift to a smaller number of latent variables.

Again, the connection to the `downwards' direction of Hume's dictum is clear: a new ontology (of latent variables) is gleaned from a set of structured data on the assumption that the variation in those data reflect a representative variation in what is `kinematically possible' for the latent variables. Cases of dimensional reduction are particularly salient, since in this case correlations between the manifest variables is explained as an artefact of having started with a misconception of the `true' numbers of degrees of freedom of the system.
\end{itemize}

To conclude, I hope to have established, or at least made plausible three main claims. First, that the philosophy of physics is in need of taking kinematical possibilities more seriously than it has done so far, since they are indispensable for theory interpretation, and more formal aims such as demonstrating inter-theoretic reductions and theoretical equivalences. Particularly if we are to take physical theories as we find them, and not according to some Quinean regimentation. Second, that these kinematical possibilities must be conceived as widely as possible, and for that we need an `upwards' Hume's dictum, according to which these possibilities are generated by a free variation of the theory's degrees of freedom. And third, that Hume's dictum may be usefully implemented in the opposite direction: `downwards' from kinematical possibilities to an ontology. Suffice it to say, I believe this third idea is worthy of much further consideration.


\section{Acknowledgements}

I am immensely grateful to Eleanor March, Erik Curiel,  Sebastian Lutz, Caspar Jacobs, Dominik Ehrenfels, Zack Glindon, James Read, Oliver Pooley, Christopher Timpson, Simon Saunders, Harvey Brown, David Wallace, Henrique Gomes, and  Jeremy Butterfield for their wisdom and generosity. 

\section{References}
\begin{itemize}
\item Abdi, H., \& Williams, L. J. (2010). Principal component analysis. Wiley interdisciplinary reviews: computational statistics, 2(4), pp. 433-459.
\item Andreas, H. (2021). Theoretical Terms in Science. The Stanford Encyclopedia of Philosophy (Fall 2021 Edition), Edward N. Zalta (ed.), URL = $\langle$https://plato.stanford.edu/archives/fall2021/entries/theoretical-terms-science/$\rangle$. 
\item Barrett, J. W. (1991). Holonomy and path structures in general relativity and yang-mills theory.
International Journal of Theoretical Physics 30, pp.~1171-1215.
\item Barrett, T. W. \& Halvorson, H. (2016). Morita equivalence. The Review of Symbolic Logic, 9(3), 556-582.
\item Beth, E. W. (1960). Semantics of Physical Theories. Synthese 12 (2-3), pp.~172-175.
\item Brown, H. R., \& Pooley, O. (2006). Minkowski space-time: A glorious non-entity. Philosophy and Foundations of Physics, 1, 67-89.
\item Butterfield, J. (2011). Against pointillisme: a call to arms. In  D. Dieks, W. Gonzalez, S. Hartmann, T. Uebel and M. Weber (eds.), Explanation, prediction, and confirmation. pp. 347-365.
\item Carnap, R. (1969). The logical structure of the world. Berkeley: University of California Press. Translation by R. George of Der logische Aufbau der Welt (Berlin: Weltkreis, 1928). 
\item Coffey, K. (2014). Theoretical Equivalence as Interpretative Equivalence. The British Journal for the Philosophy of Science, 65, 821-844.
\item Curiel, E. (2016). Kinematics, dynamics, and the structure of physical theory. Unpublished manuscript, available at: arXiv:1603.02999.
\item Dasgupta, S. (2016). Symmetry as an epistemic notion (twice over). The British Journal for the Philosophy of Science, 67, pp.~837-878.
\item Earman, J. (1989). World Enough and Space-Time. Cambridge, MA: MIT Press.
\item Earman, J. \& Friedman, M. (1973). The Meaning and Status of Newton's Law of Inertia and the Nature of Gravitational Forces. Philosophy of Science 40, pp. 329-259.
\item French, S. (2017). (Structural) realism and its representational vehicles. Synthese, 194(9), 3311-3326.
\item French, S., \& Ladyman, J. (2010). In defence of ontic structural realism. In Scientific structuralism (pp. 25-42). Dordrecht: Springer Netherlands.
\item Geroch, R. and P. S. Jang. (1975). Motion of a body in general relativity. Journal of Mathematical Physics 16, 65.
\item Goodman, N. (1977). The Structure of Appearance. 3rd ed. Dordrecht: Reidel.
\item Hudetz, L. (2019). Definable categorical equivalence. Philosophy of Science, 86(1), 47-75.

\item Hume, D. (1975). A Treatise of Human Nature. Edited by L. A. Selby-Bigge, 2nd ed. revised by P. H. Nidditch, Oxford: Clarendon Press.
\item Jacobs, C. (2023). Comparativist Theories or Conspiracy Theories: the No Miracles Argument Against Comparativism. Forthcoming in the Journal of Philosophy.
\item Ladyman, J. (1998). What is structural realism?. Studies in History and Philosophy of Science Part A, 29(3), pp.~409-424.
\item Ladyman, J. (2001). Science Metaphysics and Structural Realism. Philosophica, 67(1).
\item Ladyman, J., \& Ross, D. (2007). Every thing must go: Metaphysics naturalized. Oxford University Press.
\item Lewis, D. K. (1970). How to define theoretical terms. The Journal of Philosophy, 67(13), 427-446.
\item Lewis, D. K. (1991). Parts of Classes (with an appendix by John P. Burgess, A. P.
Hazen, and David Lewis). Basil Blackwell, Oxford.
\item Lutz, S. (2012). Criteria of empirical significance: Foundations, relations, applications. Quaestiones Infinitae, 70.
\item March, E. (2023). Are Maxwell gravitation and Newton-Cartan theory theoretically equivalent? Forthcoming in Philosophy of Science.
\item March, E. (2024). On some examples from first-order logic as motivation for categorical equivalence of KPMs. Unpublished manuscript, available at: https://philsci-archive.pitt.edu/23217/.
\item Maudlin, T. (2007). The metaphysics within physics. Oxford University Press.
\item McKenzie, K. (2017). Ontic structural realism. Philosophy Compass, 12(4), e12399.
\item  Mead, A. (1992). Review of the Development of Multidimensional Scaling Methods. Journal of the Royal Statistical Society. Series D (The Statistician), 41(1), pp.~27-39. https://doi.org/10.2307/2348634
\item Myrvold, W. C. (2019). How could relativity be anything other than physical?. Studies in History and Philosophy of Science Part B: Studies in History and Philosophy of Modern Physics, 67, 137-143.
\item Nagel, E. (1979). The Structure of Science, 2nd Edition. Hackett.
\item Pincock, C. (2009). Carnap?s logical structure of the world. Philosophy Compass, 4(6), pp.~951-961.
\item Quine, W. V. 1948. On What There Is. The Review of Metaphysics, 2, pp.~21?38.
\item Quine, W. V. 1960. Word and Object. Cambridge, MA: M.I.T.
\item Rosenstock, S., Barrett, T. W., \& Weatherall, J. O. (2015). On Einstein algebras and relativistic spacetimes. Studies in History and Philosophy of Science Part B: Studies in History and Philosophy of Modern Physics, 52, pp.~309-316.
\item Rosenstock, S., \& Weatherall, J. O. (2016). A categorical equivalence between generalized holonomy maps on a connected manifold and principal connections on bundles over that manifold. Journal of Mathematical Physics, 57(10).
\item Ruetsche, L. (2011). Interpreting quantum theories. Oxford University Press.
\item Russell, J. S., \& Hawthorne, J. (2018). Possible patterns. Oxford Studies in Metaphysics, 11, 149.
\item Rynasiewicz, R. (1992). Rings, holes and substantivalism: On the program of Leibniz algebras. Philosophy of Science, 59(4), pp.~572-589.
\item Saunders, S. (2003). Structural realism, again. Synthese, 136(1), 127-133.
\item Skyrms, B. (1993). Logical atoms and combinatorial possibility. The Journal of Philosophy, 90(5), 219-232.
\item Sneed, J. D. (1971). The Logical Structure of Mathematical Physics. Dordecht: Reidel.
\item van Fraassen, B. C. (1967). Meaning Relations Among Predicates.
No\^us 1 (2), pp.~161?179.
\item  van Fraassen, B. C. (1970). On the Extension of Beth?s Semantics of Physical Theories. Philosophy of Science 37 (3), pp.~325?339.
\item Van Fraassen, B. C. (1991). Quantum mechanics: An empiricist view. Oxford University Press.
\item Wallace, D. (2014). Deflating the Aharonov-Bohm Effect. Unpublished mauscript, available at: arXiv:1407.5073.
\item Wallace, D. (2022). Stating structural realism: mathematics?first approaches to physics and metaphysics. Philosophical Perspectives, 36(1), 345-378.
\item Weatherall, J. O. (2011). The motion of a body in Newtonian theories. Journal of Mathematical Physics, 52(3).
\item Weatherall, J. O. (2016). Are Newtonian gravitation and geometrized Newtonian gravitation theoretically equivalent?. Erkenntnis, 81, 1073-1091.
\item Weatherall, J. O. (2019a). Part 1: Theoretical equivalence in physics. Philosophy Compass, 14(5), e12592.
\item Weatherall, J. O. (2019b). Part 2: Theoretical equivalence in physics. Philosophy Compass, 14(5), e12591.
\item Weatherall, J. O. (2020). Equivalence and duality in electromagnetism. Philosophy of Science, 87(5), 1172-1183.
\item Weatherall, J. O. (2021). Why not categorical equivalence?. In Hajnal Andr\'eka and Istv\'an N\'emeti on unity of science: From computing to relativity theory through algebraic logic (pp. 427-451). Cham: Springer International Publishing.
\item Wickelmaier, F. (2003). An introduction to MDS. Sound Quality Research Unit, Aalborg University, Denmark, 46(5), pp.~1-26.
\item Wigner, E. P. (1954). Conservation laws in classical and quantum physics.
Progress of Theoretical Physics, 11, pp.~437-440.
\item Wilson, J. (2010). What is Hume's dictum, and why believe it? Philosophy and Phenomenological Research, 80(3), 595-637.
\item Wu, J., \& Weatherall, J. (2023). Between a stone and a Hausdorff space. Forthcoming in The British
Journal for the Philosophy of Science.
\item Xing, X., \& Radzihovsky, L. (2008). Nonlinear elasticity, fluctuations and heterogeneity of nematic elastomers. Annals of Physics, 323(1), pp.~105-203.
\end{itemize}

\end{document}